\documentclass[english,prl,twocolumn,superscriptaddress,letterpaper,nofootinbib]{revtex4-1}

\usepackage{ae} % or {zefonts}0
\usepackage[T1]{fontenc}
\usepackage[ansinew]{inputenc}
\usepackage{amsmath}
\usepackage{amssymb}
\usepackage{graphicx}
\usepackage{color}
\usepackage{hyperref}
\definecolor{gray}{rgb}{.5,0.5,0.5}
\definecolor{dblue}{rgb}{.2,0.2,.7}
\definecolor{ddblue}{rgb}{.1,0.1,.4}
\hypersetup{
    bookmarks=true,         % show bookmarks bar?
    unicode=false,          % non-Latin characters in Acrobat’s bookmarks
    pdftoolbar=true,        % show Acrobat’s toolbar?
    pdfmenubar=true,        % show Acrobat’s menu?
    pdffitwindow=true,      % page fit to window when opened
    pdfnewwindow=true,      % links in new window
    colorlinks=true,       % false: boxed links; true: colored links
    linkcolor=dblue,          % color of internal links
    citecolor=gray,        % color of links to bibliography
    filecolor=magenta,      % color of file links
    urlcolor=ddblue          % color of external links
}
\usepackage{epsfig}
\usepackage{booktabs}

    \def\be{\begin{eqnarray}}
    \def\ee{\end{eqnarray}}
    \def\no{\nonumber}

    \def\intl{\int\limits}

    \def\bn{\begin{enumerate}}
    \def\en{\end{enumerate}}
    \def\bi{\begin{itemize}}
    \def\ei{\end{itemize}}

    \def\({\left(\!}
    \def\){\!\right)}
    \def\<{\left\langle\,}
    \def\>{\, \right\rangle}
    \def\[{\left[}
    \def\]{\right]}

    \def\hat{\widehat}

    \def\a{\alpha}
    \def\b{\beta}

    \def\G{\Gamma}
    \def\e{\epsilon}

    \def\CO{{\cal O}}

    \def\pd{\partial}

    \def\D{{\rm D}}

  \def\mass{\textrm{m}}

\usepackage{varioref}
\usepackage{makeidx}
\makeindex

\usepackage[english]{babel}
\begin{document}

%%%%%%%%%%%%%%%%%%%%%%%%%%%%%%%%%%%%%
%%%%%%%%%%%%%%%%%%%%%%%%%%%%%%%%%%%%%
%%%%%%%%%%%%%%%%%%%%%%%%%%%%%%%%%%%%%

\title{From the mass gap in $O(N)$ to the non-Borel-summability in $O(3)$ and $O(4)$ sigma models}

\author{Dmytro Volin}
\affiliation{Institut de Physique Th\'eorique, CNRS-URA 2306, C.E.A.-Saclay, F-91191 Gif-sur-Yvette, France \&\\ Bogolyubov
Institute for Theoretical Physics, 14b Metrolohichna Str.  \\
Kyiv, 03143 Ukraine}

\begin{abstract}
We give an analytical derivation of the mass gap of the vector $O(N)$ sigma model in two dimensions and investigate a large-order behavior of the weak coupling asymptotic expansion for the energy. For sufficiently large $N$ the series is sign-oscillating, which is expected from the large $N$ solution of the sigma model. However, for $N=3$ and $N=4$ the series have positive coefficients.
\end{abstract}

\maketitle

\section{Introduction}
The two-dimensional vector $O(N)$ sigma model is often considered as a toy model for the quantum chromodynamics. It is asymptotically free and dynamically generates a mass scale $\Lambda$ (analog of $\Lambda_{QCD}$), although its classical formulation does not contain any dimensionful parameters. It is widely accepted that the asymptotic states of the sigma model are the massive particles in the vector representation of the $O(N)$ group. The mass of the particles should be proportional to the only mass scale of the theory: $\mass=c\ \Lambda$.

The mass $\mass$ is a physical quantity, while $c$ and $\Lambda$ depend on the regularization scheme. The coefficient $c$ cannot be determined from the perturbation theory. Luckily, the $O(N)$ sigma model can be studied nonperturbatively due to its integrability. The explicit expression for the coefficient $c$ in the $\overline{MS}$ scheme was found in \cite{Hasenfratz:1990ab,Hasenfratz:1990zz}:
\be\label{c}
  c=\left(\frac 8e\right)^{\frac 1{N-2}}\frac 1{\Gamma[1+\frac 1{N-2}]}.
\ee
We explain the idea of this calculation in the next section. In order to obtain $c$, one has to solve the integral equation (\ref{iequation}) in the weak-coupling regime at the leading and subleading order. At the leading order the solution was found by the application of the generalized Wiener-Hopf method (for a review of the method, see for example \cite{Forgacs:1991rs}). The subleading order was found only numerically, although with a high precision which allowed one to conjecture the expression (\ref{c}). In \cite{Balog:1992cm} it was mentioned (see reference [19] there) that J.Balog had given an analytical solution at the subleading order. However, to our knowledge, that result had not been published.

The goal of this paper is to give an analytical solution of the integral equation (\ref{iequation}) at subleading and higher orders in a recursive manner. This gives us, in particular, an analytical derivation of (\ref{c}). Also, armed with the recursive procedure, we make an estimation for the large-order behavior of the coefficients of the weak-coupling expansion and study Borel-summability properties of the model.

Another motivation for this work comes from the AdS/CFT correspondence where the appearance of the O(6) sigma model \cite{Alday:2007mf} raised the necessity of the solution of the equation (\ref{iequation}) and its generalization at next-to-subleading order \cite{Bajnok:2008it}.

\section{Integral equation}
 Let us consider the sigma model in the presence of an external field $h$ coupled to a $U(1)$ charge. Such a system was initially studied in \cite{Polyakov:1984et,Wiegmann:1985jt}. When the value of $h$ exceeds the mass gap, a finite density $\rho$ of equally polarized particles is created. At large values of $h$ the free energy of the system can be computed perturbatively due to the asymptotic freedom. Knowing the free energy density $f[h]$, we can find the energy density $\varepsilon[\rho]$ through the Legendre transform:
\be
  \varepsilon[\rho]=\textrm{min}_{h}(f[h]+\rho\, h).
\ee

It is convenient to introduce a running coupling constant $\a[\mu]$ via the relation
\be\label{al}
  \frac 1{\a}+(\Delta\!-\!1)\log\a=\log\!\[\frac{2\pi\mu\Delta}{\Lambda_{\overline{MS}}}\]\!,\;\;\; \Delta\!=\frac 1{N\!-\!2}.
\ee
The perturbative quantum field theory predicts the following expansion for the energy density:
\be
  \label{energy}
  \frac{\varepsilon[\rho]}{\rho^2}=\pi\Delta\(\a+\frac 12\a^2+\Delta\sum_{n=3}^\infty \chi_n \a^n\)+\CO\(\frac{\Lambda_{\overline{MS}}^2}{\rho^2}\)\!\!,
\ee
where $\a$ is evaluated at the scale $\mu=\rho$.

The idea of \cite{Hasenfratz:1990ab,Hasenfratz:1990zz} was that the energy of the system in the large volume can be calculated also from the asymptotic Bethe Ansatz, which explicitly contains the mass $\mass$. The energy density $\varepsilon[\rho]$ can be calculated in the thermodynamic limit, in which the number of particles $K$ and the length of the system $L$ both go to infinity with fixed $\rho=K/L$. The comparison of the result of the Bethe Ansatz and (\ref{energy}) allows one to find the coefficient c (\ref{c}).

In the thermodynamic limit the Bethe Ansatz reduces to the integral equation for the density of the rapidity distribution $\chi[\theta]$:
\be\label{iequation}
  \chi[\theta]&-&\int\limits_{-B}^{B}K[\theta-\theta']\chi[\theta']d\theta'=\mass\cosh[\theta],\;\;\theta^2<B^2;\no\\
K[\theta]&=&\frac 1{2\pi i}\frac d{d\theta}\log S_0[\theta],\\ \no
S_0[\theta]&=&-\frac{\Gamma\[\frac 12+\frac{i\theta}{2\pi}\]\Gamma\[\Delta+\frac{i\theta}{2\pi}\]}{\Gamma\[1+\frac{i\theta}{2\pi}\]\Gamma\[\frac 12+\Delta+\frac{i\theta}{2\pi}\]}/\textrm{c.c}.\ .\no
\ee
The energy density and the density of particles are given by
\be
  \varepsilon=\mass\int_{-B}^B\frac {d\theta}{2\pi}\chi[\theta]\cosh[\theta],\;\;\;\;
  \rho=\int_{-B}^B\frac {d\theta}{2\pi}\chi[\theta].
\ee
 We see that $\varepsilon$ depends on $\rho$ through the parameter $B$. To compare with the expansion (\ref{energy}) we have to consider the large $\rho$, or equivalently large $B$, asymptotics of the integral equation (\ref{iequation}).

Prior to giving a large $B$ solution of (\ref{iequation}), let us rewrite (\ref{iequation}) in terms of the resolvent for the function $\chi[\theta]$. For this we first notice that the kernel $K[\theta]$ can be represented as
\be\label{K}
  K[\theta]=
\frac 1{2\pi i} \left(\frac{\D+\D^{2 \Delta
   }}{1+\D}-\frac{\D^{-1}+\D^{-2 \Delta
   }}{1+\D^{-1}}\right)
   \frac 1{\theta},
\ee
where $\D=e^{i\pi\partial_\theta}$ is a shift operator and $(1+\D^{\pm 1})^{-1}\equiv 1-\D^{\pm 1}+\D^{\pm 2}-\ldots$. This representation for the kernel can be easily derived if to notice the formal equality $\G[-i\theta/2\pi]\simeq (1/\theta)^{\frac {\D^2}{1-\D^2}}$.

The resolvent of $\chi[\theta]$ defined by
\be\label{resolvent}
  R[\theta]=\int_{-B}^B d\theta'\frac{\chi[\theta']}{\theta-\theta'}
\ee
is analytic everywhere in the complex plane except on the support $[-B,B]$ of $\chi[\theta]$. The residue of $R[\theta]$ at infinity equals $2\pi\rho$. The density distribution $\chi[\theta]$ can be read from the discontinuity of the resolvent on the interval $[-B,B]$:
\be\label{discontinuity}
  \chi[\theta]=-\frac 1{2\pi i}(R[\theta+i0]-R[\theta-i0]).
\ee

By substituting (\ref{K}) and (\ref{discontinuity}) to (\ref{iequation}) and then evaluating the integral using (\ref{resolvent}) we obtain the following equality:
\be\label{requation}
\frac{1-\D^{2\Delta}}{1+\D}R[\theta+i0]-\frac {1-\D^{-2\Delta}}{1+\D^{-1}}R[\theta-i0]=\no\\=-2\pi i \mass\cosh[\theta],\;\; \theta^2<B^2.
\ee

\section{Leading order solution for the energy density}
In the following we will neglect the terms that give exponentially suppressed contribution to the value of $\varepsilon$. In this approximation we have
\be\label{e1}
  \frac{\varepsilon}{\mass}&\simeq&\int_{0}^B\!\! e^{\theta}\chi[\theta]\frac{d\theta}{2\pi}\simeq e^B\int_{-\infty}^0\!\!\!\! e^{\frac {z}{2}}\chi[z]\frac{dz}{4\pi},\no\\ &&\theta=B+\frac z2\;.
\ee
In other words, $\varepsilon$ receives the main contribution from the vicinity of the branch points $\pm B$. Therefore we will consider the double scaling limit
\be\label{doublescaling}
 B,\theta\to\infty,\ \ z=2(\theta-B) \ \textrm{fixed}.
\ee

From (\ref{discontinuity}) and  (\ref{e1}) we can express the energy density in terms of the inverse Laplace transform of the resolvent:
\be\label{energylaplace}
  \varepsilon=\frac {\mass\, e^B}{4\pi}\hat{R}[1/2],\;\;\; \hat{R}[s]\equiv\!\!\intl_{-i\infty+0}^{i\infty+0}\frac {dz}{2\pi i}\ e^{sz}R[z].
\ee
In the double scaling limit equation (\ref{requation}) is simplified:
\be\label{zequation}
\frac{1-\D^{2\Delta}}{1+\D}R[z+i0]-\frac {1-\D^{-2\Delta}}{1+\D^{-1}}R[z-i0]=\no\\=-\pi i \mass e^{B}e^{\frac z2},\;\; z<0.
\ee

The inverse Laplace transform of (\ref{zequation}) is derived in the appendix and is given by:
\be\label{laplaceequation}
  \sin[&2&\pi\Delta s]\! \(
   \frac{e^{i (1 - 2\Delta)\pi s}\hat{R}[s\! -\! i 0]}{\cos[\pi (s-i0)]} +
    \frac{e^{-i (1 - 2\Delta)\pi s}\hat{R}[
       s\! +\! i 0]}{\cos[\pi (s+i0)]} \)\! =\no\\&&\!\!\!\!\!= \frac {\mass}{2 i}
  e^B \(\frac 1 {s + \frac 1 2 - i 0} - \frac 1 {s + \frac 1 2 +
       i 0}\),\;\;s<0.\;\;\;\;\;\;\;\;\;\;\
\ee
To find the correct solution to (\ref{laplaceequation}) we demand the following analytical properties for $\hat R[s]$ at each order of $1/B$ expansion:
 \bi
 \item $\hat R[s]$ is analytic everywhere except on the negative real axis,
 \item $\hat R[s]$ has simple poles at $s=-n/(2\Delta)$ and simple zeroes at $s=-1/2-n$, where $n$ is a positive integer,
 \item $\hat R[s]$ is expanded in negative powers of $s$ at infinity.
 \ei
 The presence of zeroes in the resolvent can be directly seen from  equation (\ref{laplaceequation}). The solution of the correspondent homogeneous equation should have in addition a zero at $s=-1/2$.
 The origin of the other stated analytical properties is explained in the appendix.
 %The pole structure at $0$ can be very involved. T

 The most general solution of  equation (\ref{laplaceequation}) which satisfies the stated analyticity properties is the following:
 \be\label{sol1}
  \hat R[s]&=&A\,\Phi[s]\left(\frac{1}{s+\frac 12}+Q[s]\right)\!\!,\;A=\frac{\mass}{4\Delta^\Delta}e^{-\frac 12+B+\Delta}\Gamma[\Delta],\no\\
  \Phi[s]&=&\frac 1{\sqrt{s}}e^{(1-2\Delta)s\log[\frac se]-2\Delta s\log[2\Delta]}\frac{\Gamma[2\Delta s+1]}{\Gamma[s+\frac 12]},\no\\
  Q[s]&=&\frac 1{Bs} \sum_{n,m=0}^\infty\frac {Q_{n,m}[\log B]}{B^{m+n} s^{n}}.
\ee
 The dependence of $Q[s]$ on $B$ is not a consequence of (\ref{laplaceequation}) or analytical properties of the resolvent but can be deduced from the considerations of the next section. We see that $Q[s]$ contains only a finite number of terms at each order of $1/B$ expansion and therefore (\ref{sol1}) is properly defined. The leading order should be compared with the expression for $G_\pm[i\xi]$ of \cite{Forgacs:1991rs}.

 The leading order of $\varepsilon$ is given by
 \be
 \varepsilon=\frac{\mass\, A}{4\pi}\Phi[1/2]\ e^B.
 \ee

 \section{The particle density and subleading corrections}
 First note that if we apply the operator $(\D^{-1/2}+\D^{1/2})$ to the equation (\ref{requation}) we will get
 \be\label{prenondoublescaling}
  \!\!\left(\D^{-\Delta}\!-\D^{\Delta}\right)\times\hspace{12EM}\no\\ \times\left(\D^{\Delta-\frac 12}R[\theta+i0]+\D^{\frac 12-\Delta}R[\theta-i0]\right)=0.
 \ee
 The action of the shift operator is understood as analytical continuation. In particular, if $\Delta-1/2<0$ then the analytical continuation in $\D^{\Delta-\frac 12}R[\theta+i0]$ includes crossing the cut of the resolvent.

 In the previous section we found the most general solution in the double scaling limit (\ref{doublescaling}). Still we have to fix unknown coefficients $Q_{n,m}$. For this we consider a different regime. We take again $B\to\infty$ but now we will be interested in the values of the resolvent $R[\theta]$ at the distances of the order $B$ or larger from the branch points $\theta=\pm B$. In this case the shift operator can be expanded in the Taylor series $\D=1+i\pi\partial_\theta-\frac 12\pi^2(\partial_\theta)^2+\ldots$.

  Since $R[\theta]$ is an odd function as it follows from (\ref{resolvent}), it is easy to check that (\ref{prenondoublescaling}) is perturbatively equivalent to
 \be\label{nondoublescaling}
    \D^{\Delta-\frac 12}R[\theta+i0]+\D^{\frac 12-\Delta}R[\theta-i0]=0.
 \ee
 For example, if at the leading order (\ref{nondoublescaling}) is given by $X=0$ then (\ref{prenondoublescaling}) is given by $\pd_\theta X=0$ integration of which gives $X={\rm const}$. However, the constant of integration is zero due to the parity properties of the left-hand side of (\ref{nondoublescaling}).

 Solving (\ref{nondoublescaling}) perturbatively order by order we can expand the resolvent in the inverse powers of $B$:
 \be\label{sol2}
  \!\!R[\theta]\!=\!\!\!\!\sum_{n,m=0}^\infty\sum_{k=0}^{m+n}\frac{\sqrt{B}\,c_{n,m,k}(\theta/B)^{\e[k]}}{B^{m-n}\(\theta^2-B^2\)^{n+1/2}}\!\log\!\[\frac{\theta\!-\!B}{\theta\!+\!B}\]^k\!\!\!\!,
 \ee
 where $\e[k]=k\ \textrm{mod}\ 2$. The perturbative meaning of the expansion (\ref{sol2}) is most easily seen in terms of the variable $u=\theta/B$.
 %Since $R[\theta]$ should decrease at infinity, $c_{0,m,k}=0$.
 The solution (\ref{sol2}) gives us the value for the particle density from the residue of the resolvent at infinity:
 \be
  \rho=\frac{\sqrt{B}}{2\pi}\(c_{0,0,0}+\sum_{m=1}^\infty \frac{c_{0,m,0}-2c_{0,m,1}}{B^m}\).
 \ee

 If we reexpand (\ref{sol2}) in the double scaling limit (\ref{doublescaling}), we should recover the solution (\ref{sol1}) obtained in the previous section. This condition uniquely fixes all the coefficients $c_{n,m,k}$ and $Q_{n,m}$.

 The expansion (\ref{sol2}) in the double scaling limit organizes at each order of $1/B$ as a $1/z$ expansion. Therefore we should compare it with the Laplace transform of the small $s$ expansion of $\hat R[s]$. As an illustration, we give here the terms of these expansions which are relevant for calculation of the leading and the subleading orders of $\rho$ and $\varepsilon$:
 \be
  &R&=\frac {c_{0,0,0}}{\sqrt{z}}+\frac{c_{1,0,0}+c_{1,0,1}\log\[\frac{z}{4B}\]}{z^{3/2}}-\frac {\sqrt{z}}{8B} c_{0,0,0}+\no\\&+&\frac{8c_{0,1,0}-3 c_{1,0,0}-2c_{1,0,1}+(8c_{0,1,1}+c_{1,0,1})\log\[\frac{z}{4B}\]}{8B\sqrt{z}},\no\\
  &&\!\!\!\!\!\!\!\!\!\int_0^\infty\!\!\!\!ds\ e^{-sz}\hat R[s]=\frac {2A}{\sqrt z}-A\frac{2\Delta\log\!\[\frac{e\Delta}{2}\]\!\!+\!1\!+(1\!-\!2\Delta)\log z}{z^{3/2}}\no\\
  &+&\frac{Q_{0,0}\sqrt{z}}{B}\(-2+\frac{2\Delta\log\frac{2e}{\Delta}-1-(1-2\Delta)\log z}{z}\).
 \ee

\section{Results and discussions}
From the results of the previous sections we find the expressions for $\rho$ and $\varepsilon$ at the leading and subleading orders:
 \be
 \!\!\frac{\varepsilon[B]}{\mass^2}\!&=&\!\!\frac{e^{2B+2\Delta-1}}{16\pi \Delta^{2\Delta-1}}\Gamma[\Delta]^2\(1+\frac 1{4B}\),\\
 \!\!\frac{\rho[B]}{\mass}\!&=&\!\!\frac{e^{B+\Delta-\frac 12}}{4\pi\Delta^\Delta}\Gamma[\Delta]\sqrt{B}\times\no\\
 &&\times\(1-\frac {\frac 32+(1\!-\!2\Delta)\log \frac{8B}{\Delta}-\log\frac{2}{\Delta}}{4B}\).\no
 \ee
Resolving the parametric dependence we obtain exactly $\a$ and $\a^2$ terms in the expansion (\ref{energy}) if $\a$ is defined as
\be\label{am}
  \frac 1{\a}+(\Delta-1)\log \a=\log\frac{\rho}{\mass}+\log\[\(\frac{8}{e}\)^\Delta\frac{2\pi}{\Gamma[\Delta]}\].
\ee
Comparing (\ref{am}) with (\ref{al}) we confirm the result (\ref{c}).

Comparing the solutions (\ref{sol1}) and (\ref{sol2}) one can find the higher order corrections to the energy density. Up to the first four loops they are given by:
\be
&\chi_3&=\frac 12,\;\;\; \chi_4=-\frac{1}{32}  \left(24 \zeta (3) \Delta ^2
+\right.\no\\&&\left.+8
   \Delta ^2-42 \zeta (3) \Delta -28 \Delta +21 \zeta
   (3)-8\right),\\
&\chi_5&=-\frac{1}{96}\left(456 \zeta (3) \Delta
   ^3+24 \Delta ^3-918 \zeta (3) \Delta ^2\right.-\no\\&&\left.-60 \Delta
   ^2+609 \zeta (3) \Delta -140 \Delta -105 \zeta
   (3)-24\right).\no
\ee
The two-loop result $\chi_3$ coincides with the field theory calculations in \cite{Bajnok:2008it}.

The fact that the energy can be expanded in power series over the running coupling constant (\ref{am}) is a nontrivial property of the integral equation (\ref{iequation}). This is a strong check of the validity of the bootstrap approach. This hidden renorm-group dynamics of the integral equation was shown to hold in the case of Gross-Neveu models in \cite{Forgacs:1991rs}. The case of the $O(N)$ appeared to be more difficult to prove. We checked the renorm-group dynamics in the $O(N)$ case at ten first orders of the perturbative expansion and then used it to perform calculations at higher orders.

We have found analytical expressions for $\chi_n$ up to $n=26$\footnote{the \textit{Mathematica} code for the calculations is published in the PhD thesis of the author \cite{Volin:2010cq}.}. This allowed us to estimate the large $n$ behavior of $\chi_n$:
\be\label{largeorder}
  \chi_n\simeq \frac{\Gamma[n]}{2^{n-1}}a_n[\Delta].
\ee
For $\Delta=0$ we have $a_n[0]=(-1)^{n-1}$. This result is consistent with the fact that % that there is a cancellation of the IR renormalon poles in the large $N$ limit \cite{David:1982qv}. In particular this means
  in the large $N$ limit the IR renormalon poles in the Borel plane are absent \cite{David:1982qv,David:1983gz}. The large-order behavior of the coefficients $\chi_n$ is therefore governed by the leading UV renormalon pole leading to the Borel-summable oscillating behavior (\ref{largeorder}).
  %Our derivation is consistent with this prediction

   For $\Delta=1$ we have $a_n\simeq 1.09$ and for $\Delta=1/2$ we have $a_{n+1}\simeq n^{-1}(2.09 - 0.43(n\ \rm{mod}\  2))$, {\textit{i.e.}} the series is non-Borel-summable.
The Borel ambiguity is of the order ${\Lambda^2}/{\rho^2}$ as it should be from the field theory point of view (see (\ref{energy})).

For arbitrary $\Delta\geq 1$ the behavior of the coefficients $a_n$ interpolates between those for $\Delta=0$, $1/2$ and $1$. We estimate the sign oscillation of the coefficients for sufficiently large $n$ and $\Delta<\Delta_c\simeq 0.4$. For $1\leq \Delta\leq \Delta_c$ all the $a_n$ are positive. For $\Delta >1$ the asymptotic behavior of $a_n$ is given by $a_n\simeq -\a_1 n^{\a_2} \Delta^{n-2}$, where $\a_1$ is positive. This behavior shows the presence of the singularity in the Borel plane, position of which is proportional to $N-2$.

%$$\parbox[c]{4cm}{\includegraphics[width=4cm]{fig1.eps}}\parbox[c]{4cm}{\includegraphics[width=4cm]{fig2.eps}}$$

\begin{figure}[t]
\centering
\includegraphics[width=0.45\textwidth]{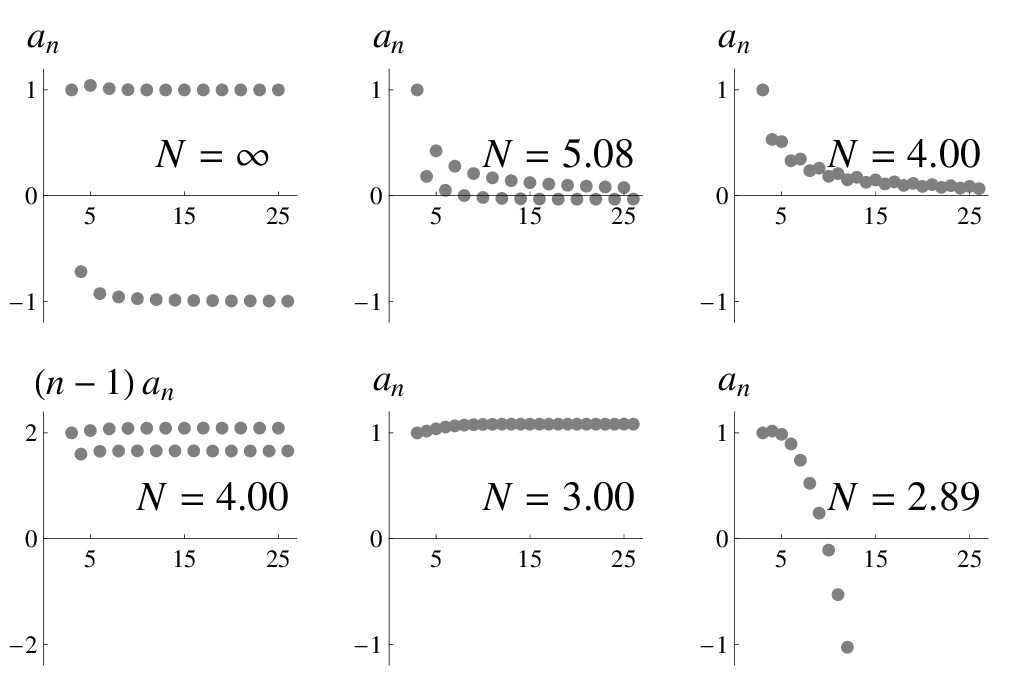}
\caption{\label{fig}Depndence of $a_n$ on $n$ for particular values of $\Delta$.}
\end{figure}

The observed non-Borel-summability in the $O(N)$ sigma models for $N=3$ and $N=4$ can be explained by the presence of IR renormalons. For $N=4$ the observation is supported by the non-Borel-summability of the $SU(N)$ principal chiral field model (PCF) at large $N$ \cite{Fateev:1994ai} (since the $O(4)$ sigma model can be viewed as the $SU(2)$ PCF model). As it follows from \cite{David:1983gz}, one should expect non-Borel-summability at any finite $N$. Although we cannot show this explicitly for arbitrary finite $N$ using argument of the positiviness of $a_n$, the observed fact that $a_n+a_{n-1}>0$ can be naturally explained by the presence of IR renormalons.

The additional singularity in the Borel plane at the positio    n proportional to $N-2$ corresponds to the instanton solution, as one can see by applying the method of \cite{Lipatov:1977hj}.

The method used in this paper was developed in \cite{Kostov:2008ax,Volin:2008kd} and
with no doubts can be applied to similar systems such as PCF or Gross-Neveu sigma models.

\section{Appendix: Analytical properties of $\hat R[s]$}
\paragraph{Expansion at infinity.} Since $K[\theta]$ is analytic on the real axis, from (\ref{iequation}) we conclude that $\chi[\theta]$ is expanded around $\theta=B$ as
\be\label{taylor}
  \chi=\b_0+(\theta-B)\b_1+(\theta-B)^2\b_2+\ldots\ .
\ee
This means that expansion of $R[z]$ at $z=0$ is  given by
\be\label{taylor2}
R[z]=-\log[z]\left(\b_0+\frac{z}2\b_1+\frac{z^2}4\b_2+\ldots\right).
\ee
The inverse Laplace transform of (\ref{taylor2}) gives the stated behavior of $\hat R[s]$ at infinity. The expansion (\ref{taylor2}) has a finite radius of convergency, therefore the inverse Laplace transform leads to a divergent asymptotic expansion. This is coherent with analytical properties of $\hat R[s]$ that we will now discuss.

\paragraph{Zeroes and poles.}
There is yet another representation of the integral equation (\ref{iequation}). To derive it we have to extend this equation to the whole real axis. For this we introduce a new function $\chi_h[\theta]$ (density of holes) which has the support complementary to $[-B,B]$. The function $\chi_h[\theta]$ is defined to be such that the equation
\be\label{iequation2}
  \chi_h[\theta']+\chi[\theta']-\int\limits_{-B}^{B}d\theta'' K[\theta'-\theta'']\chi[\theta'']=\no\\=\mass\cosh[\theta']\Theta[B^2-(\theta')^2]
\ee
is valid for any real $\theta'$.

Integrating (\ref{iequation2}) with the Cauchy kernel $\frac 1{\theta-\theta'}$, we get
\be\label{requation2}
\frac {1-\D^{\pm 2\Delta}}{1+\D^{\pm 1}}R[\theta]+R_h[\theta]&=&T[\theta],\;\;\;\; \rm{Im}[\theta]\gtrless 0,
\ee
where $R_h$ is the resolvent for $\chi_h$ and $T$ is the resolvent for the right-hand side of (\ref{iequation2}).

The function $\hat R[s]$ is defined by the inverse Laplace transform (see (\ref{energylaplace})) and is analytic for $\rm{Re}[s]>0$. We define the analytical continuation of $\hat R[s]$ to $\rm{Re}[s]<0$ using the path that does not cross the ray $s<0$. To study the analytical properties of $\hat R[s]$ in the region $\rm{Re}[s]<0,\rm{Im}[s]<0$ we consider equation (\ref{requation2}) for $\rm{Im}[\theta]>0$ in the double scaling limit and apply the integral
%\footnote{This is just an analytical continuation of the Laplace transform to the desired region. In the same way the equation (\ref{requation2}) was obtained}
$\int_{i\infty-0}^{-i\infty-0} \frac{dz}{2\pi i}e^{sz}$. The result is written as
\be\label{eq1}
  \frac{1-e^{-4i\pi\Delta s}}{1+e^{-2i\pi s}}\hat R[s]&=&\hat R_h[s]+\frac {\mass}{2}\frac {e^B}{s+\frac 12}\,,\;\; \rm{Im}[s]<0,\no\\
  \hat R_h[s]&=&\int_{i\infty-0}^{-i\infty-0} \frac{dz}{2\pi i}e^{sz} R[z].
\ee
Similarly we consider the region $\rm{Re}[s]<0,\rm{Im}[s]>0$ and get the equation
\be\label{eq2}
    \frac{1-e^{4i\pi\Delta s}}{1+e^{2i\pi s}}\hat R[s]=\hat R_h[s]+\frac {\mass}{2}\frac {e^B}{s+\frac 12}\,,\; \rm{Im}[s]>0.
\ee

Since $R_h[z]$ is analytic everywhere except on the ray $z>0$ the function $\hat R_h[s]$ is analytic for $\rm{Re}[s]<0$. Therefore, by taking the difference of (\ref{eq1}) and (\ref{eq2}) for $s<0$ we get (\ref{laplaceequation}).

We conclude from (\ref{eq1}) and (\ref{eq2}) for ${\rm Re}[s]<0$, from analyticity of $\hat R_h[s]$ for ${\rm Re}[s]<0$, and from analyticity of $\hat R[s]$ for ${\rm Re}[s]>0$  that $\hat R[s]$ is analytic everywhere except on the negative real axis where it has a cut. From (\ref{eq1}) and (\ref{eq2}) it also follows that $\hat R[s]$, on the ray $s<0$, must has zeroes at $s=n/2\Delta$ (except for the first one) and may have poles only at $s=n$. All the poles should be present in order to get a correct behavior of $\hat R[s]$ at infinity.  This explains the stated analytical structure of the resolvent.

Note that the stated analytical properties of $\hat R[s]$ are valid only in the double scaling limit. The inverse Laplace transform of $R[\theta]$ at finite $B$ is an entire function in the $s$-plane. This phenomenon is illustrated by the function $f[s]=\frac 1{s}-\frac 1{s}e^{-\frac sB}$. This is an entire function. However, since we are interested in $f[1/2]$ in the large $B$ limit the exponentially suppressed term can be neglected and effectively we get a pole.\\

{\bf Acknowledgments}

\noindent The author thanks to B. Basso, F. David, G. Korchemsky, I.Kostov, D.Serban, and J. Zinn-Justin for many useful discussions. The author is grateful to B.Basso and G.Korchemsky for providing a comprehensive introduction into the subject. This work has been supported by the  European Union
through ENRAGE network (contract MRTN-CT-2004-005616).


\begin{thebibliography}{99}
%%\cite{Hasenfratz:1990ab}
%\bibitem{Hasenfratz:1990ab}
%  \href{http://dx.doi.org/10.1016/0370-2693(90)90686-Z}{P.~Hasenfratz and F.~Niedermayer,
%  %``The Exact mass gap of the O(N) sigma model for arbitrary N is >= 3 in d =
%  %2,''
%  Phys.\ Lett.\  B {\bf 245}, 529 (1990).}
%  %%CITATION = PHLTA,B245,529;%%}
%%\cite{Hasenfratz:1990zz}
%\bibitem{Hasenfratz:1990zz}
%\href{http://dx.doi.org/10.1016/0370-2693(90)90685-Y}
%{P.~Hasenfratz, M.~Maggiore and F.~Niedermayer,
%  %``The Exact mass gap of the O(3) and O(4) nonlinear sigma models in d = 2,''
%  Phys.\ Lett.\  B {\bf 245}, 522 (1990).}
%  %%CITATION = PHLTA,B245,522;%%
%
%\bibitem{Forgacs:1991rs}
%\href{http://dx.doi.org/10.1016/0550-3213(91)90044-X}
%  {P.~Forgacs, F.~Niedermayer and P.~Weisz,
%  %``The Exact mass gap of the Gross-Neveu model. 1. The Thermodynamic Bethe
%  %ansatz,''
%  Nucl.\ Phys.\  B {\bf 367}, 123 (1991).}
%  %%CITATION = NUPHA,B367,123;%%
%
%%\cite{Balog:1992cm}
%\bibitem{Balog:1992cm}
%\href{http://dx.doi.org/10.1103/PhysRevLett.69.873}
%{J.~Balog, S.~Naik, F.~Niedermayer and P.~Weisz,
%  %``The Exact mass gap of the chiral SU(n) x SU(n) model,''
%  Phys.\ Rev.\ Lett.\  {\bf 69}, 873 (1992).}
%  %%CITATION = PRLTA,69,873;%%
%
%%\cite{Alday:2007mf}
%\bibitem{Alday:2007mf}
%\href{http://dx.doi.org/10.1088/1126-6708/2007/11/019}
%{L.~F.~Alday and J.~M.~Maldacena,
%  %``Comments on operators with large spin,''
%  JHEP {\bf 0711}, 019 (2007)}
%\href{http://arxiv.org/abs/0708.0672}
%{ [arXiv:0708.0672 [hep-th]].}
%  %%CITATION = JHEPA,0711,019;%%
%
%%\cite{Bajnok:2008it}
%\bibitem{Bajnok:2008it}
%\href{http://dx.doi.org/10.1016/j.nuclphysb.2008.11.023}
%{Z.~Bajnok, J.~Balog, B.~Basso, G.~P.~Korchemsky and L.~Palla,
%  %``Scaling function in AdS/CFT from the O(6) sigma model,''
%  Nucl.\ Phys.\  B {\bf 811}, 438 (2009)
%}
%\href{http://arxiv.org/abs/0809.4952}
%{[arXiv:0809.4952 [hep-th]].}
%  %%CITATION = NUPHA,B811,438;%%
%
%%\cite{Polyakov:1984et}
%\bibitem{Polyakov:1984et}
%\href{http://dx.doi.org/10.1016/0370-2693(84)90206-5}
%{A.~M.~Polyakov and P.~B.~Wiegmann,
%  %``Goldstone Fields In Two-Dimensions With Multivalued Actions,''
%  Phys.\ Lett.\  B {\bf 141}, 223 (1984).
%  %%CITATION = PHLTA,B141,223;%%
%}
%%\cite{Wiegmann:1985jt}
%\bibitem{Wiegmann:1985jt}
%\href{http://dx.doi.org/10.1016/0370-2693(85)91171-2}
%{P.~B.~Wiegmann,
%  %``Exact Solution Of The O(3) Nonlinear Sigma Model,''
%  Phys.\ Lett.\  B {\bf 152}, 209 (1985).
%  %%CITATION = PHLTA,B152,209;%%
%}
%
%%\cite{Volin:2010cq}
%\bibitem{Volin:2010cq}
%\href{http://dx.doi.org/10.1088/1751-8113/44/12/124003}
%{D.~Volin,
%  %``Quantum integrability and functional equations: Applications to the spectral problem of AdS/CFT and two-dimensional sigma models,''
%  J.\ Phys.\ A {\bf A44}, 124003 (2011).}
%\href{http://arxiv.org/abs/arXiv:1003.4725}
%{[arXiv:1003.4725 [hep-th]].}
%
%%\cite{David:1982qv}
%\bibitem{David:1982qv}
%\href{http://dx.doi.org/10.1016/0550-3213(82)90266-8}
%{F.~David,
%  %``Nonperturbative Effects And Infrared Renormalons Within The 1/N Expansion
%  %Of The O(N) Nonlinear Sigma Model,''
%  Nucl.\ Phys.\  B {\bf 209}, 433 (1982).
%  %%CITATION = NUPHA,B209,433;%%
%}
%%\cite{Fateev:1994ai}
%
%%\cite{David:1983gz}
%\bibitem{David:1983gz}
%\href{http://dx.doi.org/10.1016/0550-3213(84)90235-9}
%{F.~David,
%  %``On The Ambiguity Of Composite Operators, Ir Renormalons And The Status   %Of
%  %The Operator Product Expansion,''
%  Nucl.\ Phys.\  B {\bf 234}, 237 (1984).
%  %%CITATION = NUPHA,B234,237;%%
%}
%
%\bibitem{Fateev:1994ai}
%\href{http://dx.doi.org/10.1016/0550-3213(94)90405-7}
%{V.~A.~Fateev, V.~A.~Kazakov and P.~B.~Wiegmann,
%  %``Principal Chiral Field At Large N,''
%  Nucl.\ Phys.\  B {\bf 424}, 505 (1994)
%}
%\href{http://arxiv.org/abs/hep-th/9403099}
%{
%  [arXiv:hep-th/9403099].
%  %%CITATION = NUPHA,B424,505;%%
%}
%
%%\cite{Lipatov:1977hj}
%\bibitem{Lipatov:1977hj}
%\href{http://www.jetpletters.ac.ru/ps/1388/article_21069.shtml}
%{L.~N.~Lipatov,
%  %``Divergence Of Perturbation Series And Pseudoparticles,''
%  JETP Lett.\  {\bf 25}, 104 (1977)
%  [Pisma Zh.\ Eksp.\ Teor.\ Fiz.\  {\bf 25}, 116 (1977)].
%  %%CITATION = ZFPRA,25,116;%%
%}
%
%%\cite{Kostov:2008ax}
%\bibitem{Kostov:2008ax}
%\href{http://dx.doi.org/10.1088/1126-6708/2008/08/101}
%{I.~Kostov, D.~Serban and D.~Volin,
%  %``Functional BES equation,''
%  JHEP {\bf 0808}, 101 (2008)
%}
%\href{http://arxiv.org/abs/arXiv:0801.2542}
%{[arXiv:0801.2542 [hep-th]].
%  %%CITATION = JHEPA,0808,101;%%
%}
%
%%\cite{Volin:2008kd}
%\bibitem{Volin:2008kd}
%\href{http://arxiv.org/abs/arXiv:0812.4407}
%{D.~Volin,
%  %``The 2-loop generalized scaling function from the BES/FRS equation,''
%  arXiv:0812.4407 [hep-th].
%  %%CITATION = ARXIV:0812.4407;%%
%}
%
%
%%\cite{Forgacs:1991rs}
%
%
%
%
%%%\cite{Basso:2008tx}
%%\bibitem{Basso:2008tx}
%%  B.~Basso and G.~P.~Korchemsky,
%%  %``Embedding nonlinear O(6) sigma model into N=4 super-Yang-Mills theory,''
%%  Nucl.\ Phys.\  B {\bf 807}, 397 (2009)
%%  [arXiv:0805.4194 [hep-th]].
%%  %%CITATION = NUPHA,B807,397;%%
%%
%%%\cite{Beneke:1998eq}
%%\bibitem{Beneke:1998eq}
%%  M.~Beneke, V.~M.~Braun and N.~Kivel,
%%  %``The operator product expansion, non-perturbative couplings and the  Landau
%%  %pole: Lessons from the O(N) sigma model,''
%%  Phys.\ Lett.\  B {\bf 443}, 308 (1998)
%%  [arXiv:hep-ph/9809287].
%%  %%CITATION = PHLTA,B443,308;%%
%%
%%%\cite{Beneke:1998ui}
%%\bibitem{Beneke:1998ui}
%%  M.~Beneke,
%%  %``Renormalons,''
%%  Phys.\ Rept.\  {\bf 317}, 1 (1999)
%%  [arXiv:hep-ph/9807443].
%%  %%CITATION = PRPLC,317,1;%%


%\cite{Hasenfratz:1990ab}
\bibitem{Hasenfratz:1990ab}
{P.~Hasenfratz }
{and }
{F.~Niedermayer, }
  %``The Exact mass gap of the O(N) sigma model for arbitrary N is >= 3 in d =
  %2,''
\href{http://dx.doi.org/10.1016/0370-2693(90)90686-Z}
{Phys.\ Lett.\  B {\bf 245},}
\href{http://dx.doi.org/10.1016/0370-2693(90)90686-Z}
{529 (1990).}
  %%CITATION = PHLTA,B245,529;%%}
%\cite{Hasenfratz:1990zz}


\bibitem{Hasenfratz:1990zz}
{P.~Hasenfratz, M.~Maggiore and F.~Niedermayer,}
  %``The Exact mass gap of the O(3) and O(4) nonlinear sigma models in d = 2,''
\href{http://dx.doi.org/10.1016/0370-2693(90)90685-Y}
{Phys.}
\href{http://dx.doi.org/10.1016/0370-2693(90)90685-Y}
{Lett.\  B {\bf 245}, 522 (1990).}
  %%CITATION = PHLTA,B245,522;%%

\bibitem{Forgacs:1991rs}
{P.~Forgacs, F.~Niedermayer and P.~Weisz,}
  %``The Exact mass gap of the Gross-Neveu model. 1. The Thermodynamic Bethe
  %ansatz,''
\href{http://dx.doi.org/10.1016/0550-3213(91)90044-X}
{Nucl.\ Phys.\  B}
\href{http://dx.doi.org/10.1016/0550-3213(91)90044-X}
{{\bf 367}, 123 (1991).}
  %%CITATION = NUPHA,B367,123;%%

%\cite{Balog:1992cm}
\bibitem{Balog:1992cm}
{J.~Balog, S.~Naik, F.~Niedermayer and P.~Weisz,}
  %``The Exact mass gap of the chiral SU(n) x SU(n) model,''
\href{http://dx.doi.org/10.1103/PhysRevLett.69.873}
{Phys.}
\href{http://dx.doi.org/10.1103/PhysRevLett.69.873}
{Rev.\ Lett.\  {\bf 69}, 873 (1992).}
  %%CITATION = PRLTA,69,873;%%

%\cite{Alday:2007mf}
\bibitem{Alday:2007mf}
{L.~F.~Alday }
{and }
{J.~M.~Maldacena,}
  %``Comments on operators with large spin,''
\href{http://dx.doi.org/10.1088/1126-6708/2007/11/019}
{JHEP {\bf 0711}, 019}
\href{http://dx.doi.org/10.1088/1126-6708/2007/11/019}
{(2007)}
\href{http://arxiv.org/abs/0708.0672}
{[arXiv:0708.0672 [hep-th]].}
  %%CITATION = JHEPA,0711,019;%%

%\cite{Bajnok:2008it}
\bibitem{Bajnok:2008it}
{Z.~Bajnok, }
{J.~Balog, }
{B.~Basso, }
{G.~P.~Korchemsky and}
{L.~Palla,\!}
\href{http://dx.doi.org/10.1016/j.nuclphysb.2008.11.023}
{%``Scaling function in AdS/CFT from the O(6) sigma model,''
  Nucl.\ Phys. B {\bf 811}, 438\! (2009)\!\!
}
\href{http://arxiv.org/abs/0809.4952}
{[arXiv:0809.4952}
\href{http://arxiv.org/abs/0809.4952}
{[hep-th]].}
  %%CITATION = NUPHA,B811,438;%%

%\cite{Polyakov:1984et}
\bibitem{Polyakov:1984et}
{A.~M.~Polyakov }
{and }
{P.~B.~Wiegmann, }
  %``Goldstone Fields In Two-Dimensions With Multivalued Actions,''
\href{http://dx.doi.org/10.1016/0370-2693(84)90206-5}
{Phys.\ Lett.\ }
\href{http://dx.doi.org/10.1016/0370-2693(84)90206-5}
{B }
\href{http://dx.doi.org/10.1016/0370-2693(84)90206-5}
{{\bf 141},}
\href{http://dx.doi.org/10.1016/0370-2693(84)90206-5}
{223 (1984).
  %%CITATION = PHLTA,B141,223;%%
}
%\cite{Wiegmann:1985jt}
\bibitem{Wiegmann:1985jt}
{P.~B.~Wiegmann, }
  %``Exact Solution Of The O(3) Nonlinear Sigma Model,''
\href{http://dx.doi.org/10.1016/0370-2693(85)91171-2}
{Phys.\ Lett.\  B {\bf 152}, 209 (1985).
  %%CITATION = PHLTA,B152,209;%%
}
%\cite{Volin:2010cq}
\bibitem{Volin:2010cq}
{D.~Volin,\!}
  %``Quantum integrability and functional equations: Applications to the spectral problem of AdS/CFT and two-dimensional sigma models,''
\href{http://dx.doi.org/10.1088/1751-8113/44/12/124003}
{J. Phys. A\! {\bf A44},\! 124003\!}
\href{http://dx.doi.org/10.1088/1751-8113/44/12/124003}
{(2011)\!\!}
\href{http://arxiv.org/abs/arXiv:1003.4725}
{[arXiv:1003.4725 }
\href{http://arxiv.org/abs/arXiv:1003.4725}
{[hep-th]].}

%\cite{David:1982qv}
\bibitem{David:1982qv}
{F.~David,}
\href{http://dx.doi.org/10.1016/0550-3213(82)90266-8}
{%``Nonperturbative Effects And Infrared Renormalons Within The 1/N Expansion
  %Of The O(N) Nonlinear Sigma Model,''
Nucl.\ Phys.\  B {\bf 209}, 433 (1982).
  %%CITATION = NUPHA,B209,433;%%
}
%\cite{Fateev:1994ai}

%\cite{David:1983gz}
\bibitem{David:1983gz}
{F.~David,}
\href{http://dx.doi.org/10.1016/0550-3213(84)90235-9}
  %``On The Ambiguity Of Composite Operators, Ir Renormalons And The Status   %Of
  %The Operator Product Expansion,''
{Nucl.\ Phys.\  B {\bf 234}, 237 (1984).
  %%CITATION = NUPHA,B234,237;%%
}

\bibitem{Fateev:1994ai}
{V.~A.~Fateev, V.~A.~Kazakov and P.~B.~Wiegmann, }
\href{http://dx.doi.org/10.1016/0550-3213(94)90405-7}
{Nucl.}
%\href{http://dx.doi.org/10.1016/0550-3213(94)90405-7}
%{V.~A.~Kazakov }
%\href{http://dx.doi.org/10.1016/0550-3213(94)90405-7}
%{and}
%\href{http://dx.doi.org/10.1016/0550-3213(94)90405-7}
%{P.~B.~Wiegmann, }
%\href{http://dx.doi.org/10.1016/0550-3213(94)90405-7}
%{%``Principal Chiral Field At Large N,''
%Nucl.\ }
\href{http://dx.doi.org/10.1016/0550-3213(94)90405-7}
{Phys.\  B {\bf 424}, 505 (1994)
}
\href{http://arxiv.org/abs/hep-th/9403099}
{[arXiv:hep-th/9403099].
  %%CITATION = NUPHA,B424,505;%%
}

%\cite{Lipatov:1977hj}
\bibitem{Lipatov:1977hj}
{L.~N.~Lipatov, }
  %``Divergence Of Perturbation Series And Pseudoparticles,''
\href{http://www.jetpletters.ac.ru/ps/1388/article_21069.shtml}
{JETP Lett.\ }
\href{http://www.jetpletters.ac.ru/ps/1388/article_21069.shtml}
{{\bf 25}, }
\href{http://www.jetpletters.ac.ru/ps/1388/article_21069.shtml}
{104 (1977) }
\href{http://www.jetpletters.ac.ru/ps/1388/article_21069.shtml}
{[Pisma Zh.}
\href{http://www.jetpletters.ac.ru/ps/1388/article_21069.shtml}
{Eksp.\ Teor.\ Fiz.\  {\bf 25}, 116 (1977)].
  %%CITATION = ZFPRA,25,116;%%
}

%\cite{Kostov:2008ax}
\bibitem{Kostov:2008ax}
{I.~Kostov, }
{D.~Serban }
{and }
{D.~Volin,}
  %``Functional BES equation,''
\href{http://dx.doi.org/10.1088/1126-6708/2008/08/101}
{JHEP {\bf 0808}, 101}
\href{http://dx.doi.org/10.1088/1126-6708/2008/08/101}
{(2008)
}
\href{http://arxiv.org/abs/arXiv:0801.2542}
{[arXiv:0801.2542 [hep-th]].
  %%CITATION = JHEPA,0808,101;%%
}

%\cite{Volin:2008kd}
\bibitem{Volin:2008kd}
{D.~Volin,}
  %``The 2-loop generalized scaling function from the BES/FRS equation,''
\href{http://arxiv.org/abs/arXiv:0812.4407}
{arXiv:0812.4407 [hep-th].
  %%CITATION = ARXIV:0812.4407;%%
}


%\cite{Forgacs:1991rs}




%%\cite{Basso:2008tx}
%\bibitem{Basso:2008tx}
%  B.~Basso and G.~P.~Korchemsky,
%  %``Embedding nonlinear O(6) sigma model into N=4 super-Yang-Mills theory,''
%  Nucl.\ Phys.\  B {\bf 807}, 397 (2009)
%  [arXiv:0805.4194 [hep-th]].
%  %%CITATION = NUPHA,B807,397;%%
%
%%\cite{Beneke:1998eq}
%\bibitem{Beneke:1998eq}
%  M.~Beneke, V.~M.~Braun and N.~Kivel,
%  %``The operator product expansion, non-perturbative couplings and the  Landau
%  %pole: Lessons from the O(N) sigma model,''
%  Phys.\ Lett.\  B {\bf 443}, 308 (1998)
%  [arXiv:hep-ph/9809287].
%  %%CITATION = PHLTA,B443,308;%%
%
%%\cite{Beneke:1998ui}
%\bibitem{Beneke:1998ui}
%  M.~Beneke,
%  %``Renormalons,''
%  Phys.\ Rept.\  {\bf 317}, 1 (1999)
%  [arXiv:hep-ph/9807443].
%  %%CITATION = PRPLC,317,1;%%

\end{thebibliography}
\end{document}